**Magnetized Rutherford scattering angle for electron-ion collision in plasma**

Chang Jiang, [1, 2] Ding Li, [1, 3, 2, a] and Chao Dong[1, 2]

**AFFILIATIONS**

[1]Laboratory of Soft Matter Physics, Institute of Physics, Chinese Academy of Sciences, Beijing 100190, China

[2]University of Chinese Academy of Sciences, Beijing 100049, China

[3]Songshan Lake Materials Laboratory, Dongguan, Guangdong 523808, China

[a]Author to whom correspondence should be addressed: dli@iphy.ac.cn

**ABSTRACT**

Rutherford scattering formula plays an important role in plasma classical transport. It is urgent to need a magnetized Rutherford scattering formula since the magnetic field increases significantly in different fusion areas (e.g. tokamak magnetic field, self-generated magnetic field, and compressed magnetic field). The electron-ion Coulomb collisions perpendicular to the external magnetic field are studied in this paper. The scattering angle is defined according to the electron trajectory and asymptotic line (without magnetic field). A magnetized Rutherford scattering formula is obtained analytically under the weak magnetic field approximation. It is found that the scattering angle decreases as external magnetic field increases. It is easy to find the scattering angle decreasing significantly as incident distance, and incident velocity increasing. It is shown that the theoretical results agree well with numerical calculation by checking the dependence of scattering angle on external magnetic field.

## I. INTRODUCTION

The Coulomb collision process is the basis for plasma classical transport. Collisional process should be dominated by binary scattering events in plasma. The multi particle interaction in plasma can be approximately regarded as the superposition of a series of independent binary collisions.

Rutherford found the scattering formula of charged particles in 1911.[1] The scattering also helped him found the structure of atom. Rutherford scattering formula is for central force fields in which external magnetic field has not been included. M. M. Gordon concerned the importance of inverse-square-law in Rutherford scattering formula in 1955.错误!未找到引用源。

A. D. Fokker[3] and M. Plank[4] derived Fokker-Planck collision term respectively in 1914 and 1917. L. Landau and E. Teller used Rutherford scattering formula to derive Landau collision term in 1936.[5] Landau collision term was cut off by the Debye Length and Landau Length in plasma physics. Early fusion devices in 1950s had a relative weak magnetic field, in such case Larmor radius of the electron ($\rho_e$) is larger than Debye Length ($\lambda_D$). Therefore, an electron won't rotate much in a binary collision process. The collision processes can be solved analytically without considering an external magnetic field. M. N. Rosenbluth, W. M. Macdonald, and D. L. Judd derived Fokker-Planck equation for an arbitrary distribution function in 1957.[6] In 1965, E. R. Wicher built a Rutherford scattering simulator to introduce the scattering experiment in laboratory.[7] Chang and Li used small-momentum-transfer to replace the small-angle as cutoff variable for Fokker-Planck equation in 1996.[8]

It is necessary to consider the influence of the external magnetic field on the binary collisional scattering since the magnetic field in fusion devices increases, such as the toroidal magnetic field in magnetic confinement fusion, self-generated magnetic field in initial confinement fusion and compressed magnetic field in magnetic-initial fusion. For example, the toroidal magnetic field is 3.5T, 5.3T, 6.5T, and 12T respectively for Experimental Advanced Superconducting Tokamak (EAST), International Thermonuclear Experimental Reactor (ITER), China Fusion Engineering Test Reactor (CFETR) and SPARC tokamak. It is urgent to need a magnetized Rutherford scattering formula since the binary collisional scattering may significantly affect the plasma transport, heating, and magnetic confinement.

For example, electron Larmor radius $\rho_e$ can be smaller than the Debye Length $\lambda_D$ in plasma as the magnetic field in fusion devices increasing. The electron rotation in magnetic field can play an important role in a scattering process. It is necessary to concern binary collisions with external magnetic fields in different conditions. There are two ways to include an external magnetic field in the scattering process.

One is to derive a collision term with external magnetic field. Matsuda analyzed both electron-electron collisions and electron-ion collisions in a strong magnetic field in 1982 and 1983.[9,10] She used the Rostoker formula[11] to compute the friction and diffusion experienced by a test electron due to interactions with a Maxwellian ion distribution. She found the effect prominent when the magnetic field is strong. O' Neil has considered the extreme case for very cold electron plasmas with a strong magnetic field.[12] He found that for electron-electron collisions the close encounters are the dominant effect. In 1989, A. A. Ware has considered a case with strong magnetic field and derived a form of Fokker-Planck equation with $b > \rho_e$, where $b$ is the impact parameter.[13] Psimopoulos and Li constructed a non-local kinetic equation for a plasma in a very strong magnetic field.[14] They found that the cross-field heat transport occurs even though mass transport only along the field lines. Christian Toepffer has studied electron-ion collisions with second-order perturbation theory in 2002.[15] He found that the magnetic field suppresses the velocity transfer in the transverse direction, but it enhances the longitudinal velocity transfer. In 2008, F. L. Hinton concerned ion-ion collisions in magnetic field with distribution function independent of the gyrophase angle which would be used in simulation as a part of PIC (Particle In Cell) code.[16] C. Dong, H. Ren, H. Cai, and D. Li found that the magnetic field greatly affects the plasma's relaxation processes.[17] They obtained the expressions for the time rates of change of the electron and ion parallel and perpendicular temperatures. M. Honda derived the Coulomb logarithm formulae for collisions between particles with different temperatures in 2013.[18] C. Dong, W. Zhang, and D. Li derived a general form of the Fokker-Planck collision terms including an external magnetic field by using perturbation theory in uniform plasma in 2016.[19] In 2017, C. Dong, W. Zhang, J. Cao, and D. Li derived the Balescu-Lenard-Guernsey collision term including an external magnetic field by using wave theory.[20]

Another way to solve the magnetized scattering is to derive a scattering formula including an external magnetic field. Some authors tried to find the scattering angle in simulation ways. For instance, J. G. Siambis solved like-particle collisions numerically and found the reduced-mass and the center-of-mass motion are decoupled so that the cases are same for with/without external magnetic field in 1976.[21] D. K. Geller and J. C. Weisheit treated the scattering as a perturbation of the helical motion of the electron. For classical, small angle, electron-ion scattering, he derived a generalized Coulomb logarithm in 1997 and 1998.[22] They found that the generalized Coulomb logarithm is nearly independent of electron pitch angle. In 2000, S. A. Koryagin has considered a case with strong magnetic field.[23] He defined a differential scattering cross section with the pitch angle $\theta$, the initial rotation phase $\varphi_0$, and the modulus of the velocity $v$. He found a general integral form of both the electron-ion and electron-electron collision in a magnetic field in Boltzmann's form. Then he found only close collisions determine the form of the electron–ion collision integral in strong magnetic fields. For large angle scattering events, B. Hu, W. Horton, C. Chiu, and T. Petrosky showed complex chaotic scattering interaction events for low-energy electrons. The scattering angle has a fractal dependence on the impact parameter in the chaotic scattering intervals in 2002.[24]

So far, it seems that no one has got a satisfactory Rutherford formula with magnetic field, including scattering angle and cross section. The present paper will concern electron-ion collisions in a two-dimensional case. We will start with Newtonian mechanics to analyze binary collisional scattering with magnetic field. We are trying to derive a formula for scattering angle with external magnetic field.

The electron trajectory (with magnetic field) and its asymptote (in the case without magnetic field) will cross each other. It may be reasonable to define the scattering angle as the angle between the initial velocity direction of the electron and the radius vector direction of the cross point.

Then, we need to derive a trajectory equation and asymptote equation analytically. But we found it hard to get an analytical solution for such complex trajectory equation. We can derive the scattering angle formula from the conservation of energy and the conservation of angular momentum for the case with a relatively weak magnetic field.

The paper is organized as follows. In Section II, we will define the scattering angle in magnetic field and derive the asymptotic line equation. In Section III, we will derive the trajectory equation of the electron with magnetic field. In Section IV, we will result in the cross point of asymptote and the electron trajectory and derive a simple formula for scattering angle in relatively weak magnetic field. It is shown that the approximation of theoretical results is reasonable by comparing with numerical calculations. In Section V, the conclusions and discussions will be given.

## II. DEFINITION OF THE SCATTERING ANGLE AND THE ASYMPTOTE EQUATION

In Rutherford scattering, the scattering angle depends on the change of relative velocity before and after collision. Considering the electron-ion collision and assuming that the ion is stationary, the trajectory of the incident electron under the Coulomb force is hyperbolic. Therefore, the scattering angle is actually the angle between the direction of the initial velocity of the electron and the direction of the asymptote of the trajectory after the collision. Under the influence of external magnetic field, the trajectory of electron depends on the superposition of Larmor precession and Coulomb interaction. Under the influence of the external magnetic field, the trajectory of electrons will be superimposed on the basis of Rutherford scattering by a Larmor cyclotron process. The electron does not always move in the direction of asymptote after collision. Therefore, how to define the scattering angle becomes a difficult problem for us when there is an external magnetic field.

The electron has a complex trajectory in a magnetized collision process. The scattering process will be confined to the plane of the collision if we can consider a two-dimensional electron-ion collision with a magnetic field $B$ perpendicular to the incident plane. The problem will be reduced to the much simpler case where the physical picture of Coulomb collision is only more complex a little bit than the case without magnetic field.

When the deflection caused by cyclotron motion offsets partially the Rutherford scattering angle, the trajectory of the electron will form a cross point with the asymptote defined from the case without magnetic field. It is a reasonable choice to use the position of the cross point in the collision center coordinate to define the new scattering angle with the external magnetic field.

The scattering angle is defined as the angle between the direction of the initial velocity of the electron and the direction of the radius vector of the cross point.

The electron trajectory has no intersection with the asymptote in the absence of an external magnetic field if the deflection caused by the cyclotron motion increases the Rutherford scattering angle. In this case, the scattering angle cannot be defined.

In the weak magnetic field, the cross point is far from the collision center so that it seems to be a good approximation to take the cross point as the mark of the end of the collision process. At the same time, when the external magnetic field approaches zero, it is easy to see that the scattering angle defined in this way can return to the case without magnetic field. In other words, the result derived in the weak magnetic field limit can recover the conclusion of the Rutherford scattering.

On the other hand, we can't assume that an electron is incident from infinity in a motion with the external magnetic field. We need to calculate the electron trajectory and its asymptote when the electron incidents from a finite distance. As shown in Figure 1, like Rutherford scattering, we first establish a polar coordinate system with proton at the center and the electron anti-incident direction as the polar axis. Suppose that the distance between the incident electron and the proton is $r_0$, the incident velocity is $v_0$, the electron charge is $-e$, the mass is $m$ and the collision parameter is $b$. Let's first calculate the trajectory and asymptote of the electron in the absence of an external magnetic field.

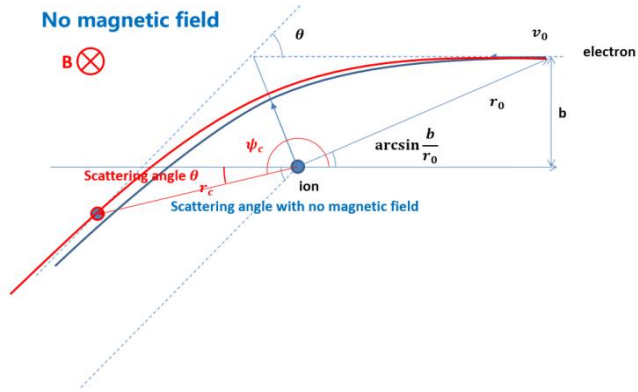

**FIG. 1.** The diagram of electron scattering by ions shows the trajectory (the blue one without magnetic field) and its asymptote which intersects with the trajectory (the red one with magnetic field) at $(r_c, \psi_c)$.

We can write the equations for the conservation of energy and the conservation of angular momentum in the International System of Units by using the polar coordinates $(r, \ \psi)$ as following:

$$\frac{1}{2}mv_0^2 - \frac{e^2}{4\pi\varepsilon_0 r_0} = \frac{1}{2}mv^2 - \frac{e^2}{4\pi\varepsilon_0 r} \tag{1}$$

$$mv_0 b = mr^2 \frac{d\psi}{dt}, \tag{2}$$

where $m$ and $-e$ is respectively the electron mass and electron charge, $v_0$, $r_0$ and $v$ is initial distance, initial velocity and velocity of the incident electron respectively, and $b$ is collision parameter. It is easy to get the differential equation for the outgoing trajectory by using the similar way from the deriving of the Rutherford scattering formula:

$$\frac{d\psi}{dr} = \frac{b}{r^2\sqrt{1 + \frac{2b_0}{r_0} - \frac{2b_0}{r} - \frac{b^2}{r^2}}}, \tag{3}$$

where $b_0 = -e^2/4\pi\varepsilon_0 mv_0^2$ is the minimum collision parameter which is negative because the electron experiences an attraction field of ion. $d\psi/dr$ is negative for the incident trajectory equation. We can ubstitute the initial condition

$$\psi = \arcsin\frac{b}{r_0} \ \ at \ r = r_0 \tag{4}$$

into the incident trajectory equation. By combining the incident and outgoing equation with the closest point, we derive the outgoing trajectory equation as:

$$\psi = C - \arcsin \frac{\frac{b^2}{r} + b_0}{\sqrt{b_0^2 + b^2 \left(1 + \frac{2b_0}{r_0}\right)}} \qquad (5)$$

where

$$C = \arcsin \frac{b}{r_0} + \pi - \arcsin \frac{\frac{b^2}{r_0} + b_0}{\sqrt{b_0^2 + b^2 \left(1 + \frac{2b_0}{r_0}\right)}}. \qquad (6)$$

We can derive the asymptote equation analytically from the trajectory equation of the incident electron as

$$\psi = \arcsin \frac{b}{r_0} + \pi - \arcsin \frac{\frac{b^2}{r_0} + b_0}{\sqrt{b_0^2 + b^2 \left(1 + \frac{2b_0}{r_0}\right)}} - \arcsin \frac{b}{r \sqrt{1 + \frac{2b_0}{r_0}}} - \arcsin \frac{b_0}{\sqrt{b_0^2 + b^2 \left(1 + \frac{2b_0}{r_0}\right)}}. \quad (7)$$

The asymptote equation is obtained under the condition without magnetic field. In a Coulomb collision with the external magnetic field, as mentioned above, the electron trajectory will not continue along the asymptote defined from the case without magnetic field due to the superposition of the electron cyclotron motion whereas the electron trajectory will intersect with this asymptote. This asymptote will be used as a reference for us to calculate the scattering angle.

## III. THE ELECTRON TRAJECTORY EQUATION IN MAGNETIC FIELD

We still use the previous polar coordinate system and incident parameters to calculate the electron trajectory after adding external magnetic field.

We have the Newton mechanic equation in the external magnetic field

$$-e\mathbf{E} - e\mathbf{v} \times \mathbf{B} = m \frac{d^2\boldsymbol{r}}{dt^2}, \qquad (8)$$

where $\mathbf{E} = e/4\pi\varepsilon_0 r^2$ and $\mathbf{B}$ is respectively electric field and magnetic field. In a polar coordinate, we define $\mathbf{e}_r$ and $\mathbf{e}_\psi$ as the unit vector. Equation (9) can be written as

$$\left(-\frac{e^2}{4\pi\varepsilon_0 r^2} + er\frac{d\psi}{dt}B\right)\mathbf{e}_r - e\frac{dr}{dt}B\mathbf{e}_\psi = m\left[\frac{d^2r}{dt^2} - r\left(\frac{d\psi}{dt}\right)^2\right]\mathbf{e}_r + m\left(2\frac{dr}{dt}\frac{d\psi}{dt} + r\frac{d^2\psi}{dt^2}\right)\mathbf{e}_\psi. \qquad (9)$$

The equation (9) can be written respectively in $\mathbf{e}_r$ and $\mathbf{e}_\psi$ directions as

$$-\frac{e^2}{4\pi\varepsilon_0 r^2} + er\frac{d\psi}{dt}B = m\left[\frac{d^2r}{dt^2} - r\left(\frac{d\psi}{dt}\right)^2\right], \qquad (10)$$

$$-e\frac{dr}{dt}B = m\left(2\frac{dr}{dt}\frac{d\psi}{dt} + r\frac{d^2\psi}{dt^2}\right). \qquad (11)$$

We can integrate Eq. (11) over the time $t$ to obtain

$$r^2\left(\frac{d\psi}{dt} + \frac{eB}{2m}\right) = C_1 \qquad (12)$$

where $C_1$ is a constant. From the initial condition

$$r_0^2 \frac{d\psi_0}{dt} = v_0 b,$$

we have

$$C_1 = r_0^2 \left( \frac{d\psi_0}{dt} + \frac{eB}{2m} \right) = v_0 b + \frac{eBr_0^2}{2m}$$

so that equation (12) becomes

$$\frac{d\psi}{dt} = -\frac{eB}{2m} + \frac{1}{r^2} \left( v_0 b + \frac{eBr_0^2}{2m} \right). \tag{13}$$

Because Lorentz force does not contribute to the energy exchange, the energy conservation equation is

$$\frac{1}{2}mv_0^2 - \frac{e^2}{4\pi\varepsilon_0 r_0} = \frac{1}{2}mv^2 - \frac{e^2}{4\pi\varepsilon_0 r}, \tag{14}$$

where

$$v = \sqrt{\left( \frac{dr}{dt} \right)^2 + r^2 \left( \frac{d\psi}{dt} \right)^2}. \tag{15}$$

Substituting Eq. (15) into Eq. (14), we have

$$mv_0^2 - \frac{2e^2}{4\pi\varepsilon_0 r_0} + \frac{2e^2}{4\pi\varepsilon_0 r} = m\left[ \left( \frac{dr}{dt} \right)^2 + r^2 \left( \frac{d\psi}{dt} \right)^2 \right]. \tag{16}$$

The expression of $dr/dt$ is sorted out as

$$\left( \frac{dr}{dt} \right)^2 = v_0^2 - \frac{2e^2}{4\pi\varepsilon_0 mr_0} + \frac{2e^2}{4\pi\varepsilon_0 mr} - r^2 \left( \frac{d\psi}{dt} \right)^2. \tag{17}$$

Substituting Eq. (13) into Eq. (17), Equation (17) becomes

$$\left( \frac{dr}{dt} \right)^2 = v_0^2 - \frac{2e^2}{4\pi\varepsilon_0 mr_0} + \frac{2e^2}{4\pi\varepsilon_0 mr} - r^2 \left[ -\frac{eB}{2m} + \frac{1}{r^2} \left( v_0 b + \frac{eBr_0^2}{2m} \right) \right]^2. \tag{18}$$

Combining Eq. (13) and Eq. (18), we get the differential form of the trajectory equation

$$\frac{d\psi}{dr} = \frac{\frac{\omega}{2v_0} + \frac{b}{r^2} - \frac{\omega r_0^2}{2v_0 r^2}}{\sqrt{1 + \frac{2b_0}{r_0} - \frac{2b_0}{r} - r^2 \left( \frac{\omega}{2v_0} + \frac{b}{r^2} - \frac{\omega r_0^2}{2v_0 r^2} \right)^2}} \tag{19}$$

where $\omega$ is defined as：

$$\omega = -\frac{eB}{m}.$$

Here we derived a differential equation (19) for the trajectory of an electron. However, the denominator of the equation contains a quartic polynomial in the square root, which is too complicate to be solved. Considering that our physical picture itself is more suitable for the weak magnetic field, we will directly adopt the weak magnetic field approximation here. In the weak magnetic field approximation, we can get some simple results.

We defined the position of the cross point as $(\psi_c, r_c)$. For weaker magnetic fields, we first assume that

$$\frac{\omega r_0^2}{bv_0} = \epsilon. \tag{20}$$

With this approximation，Equation (19) becomes a simpler form as

$$\frac{d\psi}{dr} \approx \frac{\left(1-\frac{\epsilon}{2}\right)\frac{b}{r^2}}{\sqrt{\left(1+\frac{2b_0}{r_0}-\epsilon\frac{b^2}{r_0^2}\right)-\frac{2b_0}{r}-\frac{b^2}{r^2}(1-\epsilon)}}$$

$$+\frac{\frac{\omega}{2v_0}}{\sqrt{\left(1+\frac{2b_0}{r_0}-\epsilon\frac{b^2}{r_0^2}\right)-\frac{2b_0}{r}-\frac{b^2}{r^2}(1-\epsilon)}}. \tag{21}$$

Integrating this equation over r, we can obtain the trajectory equation until $\epsilon$ order

$$\psi = \arcsin\frac{-b^2(1-\epsilon)\frac{1}{r}-b_0}{\sqrt{\left(1+\frac{2b_0}{r_0}\right)b^2(1-\epsilon)+b_0^2}} + \frac{b}{2r_0^2}\frac{\sqrt{\left(1+\frac{2b_0}{r_0}\right)r^2-2b_0r-b^2(1-\epsilon)}}{\left(1+\frac{2b_0}{r_0}\right)}\epsilon + C_2.$$

where $C_2$ is a constant. With the initial condition

$$\psi = \arcsin\frac{b}{r_0} \ at \ r=r_0,$$

we have

$$C_2 = \pi + \arcsin\frac{b}{r_0} - \arcsin\frac{b^2(1-\epsilon)\frac{1}{r_0}+b_0}{\sqrt{\left(1+\frac{2b_0}{r_0}\right)b^2(1-\epsilon)+b_0^2}} + \frac{b\sqrt{r_0^2-b^2}}{2r_0^2\left(1+\frac{2b_0}{r_0}\right)}\epsilon$$

So that the outgoing trajectory equation until $\epsilon$ order is obtained as

$$\psi = \arcsin\frac{-b^2(1-\epsilon)\frac{1}{r}-b_0}{\sqrt{\left(1+\frac{2b_0}{r_0}\right)b^2(1-\epsilon)+b_0^2}} + \frac{b\sqrt{\left(1+\frac{2b_0}{r_0}\right)r^2-2b_0r-b^2(1-\epsilon)}}{2r_0^2\left(1+\frac{2b_0}{r_0}\right)}\epsilon + \pi$$

$$+ \arcsin\frac{b}{r_0} - \arcsin\frac{b^2(1-\epsilon)\frac{1}{r_0}+b_0}{\sqrt{\left(1+\frac{2b_0}{r_0}\right)b^2(1-\epsilon)+b_0^2}} + \frac{b\sqrt{r_0^2-b^2}}{2r_0^2\left(1+\frac{2b_0}{r_0}\right)}\epsilon. \tag{22}$$

## IV. THE CROSS-POINT AND SCATTERING ANGLE IN MAGNETIC FIELD

The coordinates of the intersection point $(r_c, \psi_c)$ can be obtained by solving the simplified trajectory equation and the asymptote equation. Because both the incident distance and the distance between intersection point and center are much larger than the collision parameter, we add the hypothesis conditions in the derivation process for simplicity as following:

$$\frac{b^2}{r_c^2} \sim \frac{b_0^2}{r_c^2} \sim \epsilon, \qquad \frac{b^2}{r_0^2} \sim \frac{b_0^2}{r_0^2} \sim \epsilon. \tag{23}$$

We can derive the expression for cross point coordinates $r_c$ and $\psi_c$. If only keeping the first order term of $\epsilon$, we have the simplified result by combining Eqs. (7) and (22)

$$\arcsin\frac{b^2\frac{1}{r_c}+b_0}{\sqrt{\left(1+\frac{2b_0}{r_0}\right)b^2+b_0^2}} - \arcsin\frac{b}{r_c\sqrt{1+\frac{2b_0}{r_0}}} + \frac{b_0 b}{(b^2+b_0^2)}\epsilon = \arcsin\frac{b_0}{\sqrt{b_0^2+b^2\left(1+\frac{2b_0}{r_0}\right)}}. \tag{24}$$

Combining the first and second terms in left hand side (LHS) of Eq. (24), we found the zero order terms offset each other. We have

$$\frac{1}{r_c^2} \approx -\frac{2}{b^2 + b_0^2}\epsilon.$$ (25)

From Eq. (25), we can obtain the expression of $r_c$ as

$$r_c = \sqrt{-\frac{b^2 + b_0^2}{2\epsilon}} = \sqrt{\frac{(b^2 + b_0^2)bmv_0}{2eBr_0^2}}$$ (26)

Substituting the Eq. (26) into Eq. (7), we can get the expression of $\psi_c$ as

$$\psi_c = \arcsin\frac{b}{r_0} + \pi - \arcsin\frac{\frac{b^2}{r_0} + b_0}{\sqrt{b_0^2 + b^2\left(1 + \frac{2b_0}{r_0}\right)}}$$

$$-\arcsin\frac{b_0}{\sqrt{b_0^2 + b^2\left(1 + \frac{2b_0}{r_0}\right)}} - \arcsin\left[\frac{b}{\sqrt{(b^2 + b_0^2)\left(1 + \frac{2b_0}{r_0}\right)}}\sqrt{\frac{2eBr_0^2}{mv_0b}}\right].$$ (27)

Based on the definition of scattering angle with external magnetic field, it is easy to find the relation between azimuth angle $\psi_c$ and the scattering angle $\theta$ as

$$\psi_c - \theta = \pi.$$

Therefore, the scattering angle $\theta$ can be expressed as

$$\theta = \arcsin\frac{b}{r_0} - \arcsin\frac{\frac{b^2}{r_0} + b_0}{\sqrt{b_0^2 + b^2\left(1 + \frac{2b_0}{r_0}\right)}}$$

$$-\arcsin\frac{b_0}{\sqrt{b_0^2 + b^2\left(1 + \frac{2b_0}{r_0}\right)}} - \arcsin\left[\frac{b}{\sqrt{(b^2 + b_0^2)\left(1 + \frac{2b_0}{r_0}\right)}}\sqrt{\frac{2eBr_0^2}{mv_0b}}\right].$$ (28)

Combining the first and second terms in right hand side (RHS) of Eq. (28), and making perturbation expansion until $\epsilon$ order, equation (28) becomes:

$$\theta \approx -\arcsin\frac{2b_0 b\sqrt{\left(1 + \frac{2b_0}{r_0}\right)}}{b_0^2 + b^2\left(1 + \frac{2b_0}{r_0}\right)}$$

$$-\arcsin\left[\frac{b}{\sqrt{(b^2 + b_0^2)\left(1 + \frac{2b_0}{r_0}\right)}}\sqrt{\frac{2eBr_0^2}{mv_0b}}\right] + \frac{\sqrt{b_0^2 + b^2\left(1 + \frac{2b_0}{r_0}\right)}}{1 + \frac{2b_0}{r_0}}\frac{b}{2r_0^2}.$$ (29)

The second term on the RHS of the scattering angle formula is due to the influence of magnetic field, and the first and the third terms include the influence of the incident position until $\epsilon$ order, which was absent in the original Rutherford formula. If the effects of magnetic field and incident position are ignored, equation (29) will be reduced to the classical formula of Rutherford scattering angle as

$$\theta \approx -\arcsin \frac{2b_0 b}{b_0^2 + b^2} \tag{30}$$

Here $b_0$ is a negative value, representing the attractive field.

We will analyze the influence of the magnetic field and the incident position on scattering angle respectively.

Firstly, the effect of external magnetic field on scattering angle is considered. The derivative of Eq. (29) for magnetic field can be obtained as

$$\frac{d\theta}{dB} \approx -\frac{er_0^2}{mv_0\sqrt{(b^2+b_0^2)\left(1+\frac{2b_0}{r_0}\right)-\frac{2bBer_0^2}{mv_0}}\sqrt{\frac{2eBr_0^2}{mv_0 b}}} < 0 \tag{31}$$

Obviously, the increase of magnetic field will lead to the decrease of the scattering angle in our case. As shown in Fig. 1, the deflection direction caused by the selected magnetic field is opposite to that caused by scattering. Therefore, the enhancement of magnetic field results in the reduction of the scattering angle.

As shown in Fig. 2, the scattering angle decreases significantly as magnetic field increases for different incident distance and $b = 3 * 10^{-12}\,\mathrm{m}, v_0 = 1.59 * 10^7\,\mathrm{m/s}$ （corresponding to $b_0 = 1 * 10^{-12} m$, $T = 1.11 * 10^7\mathrm{K}$）

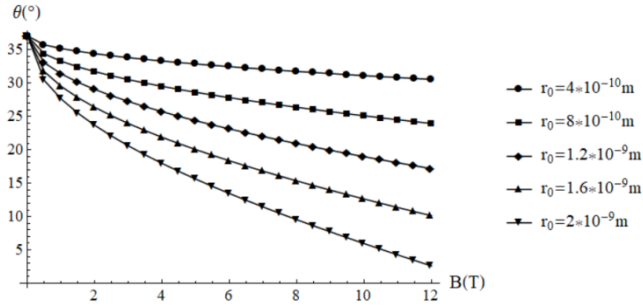

FIG. 2. The scattering angle decreases as B increases in different $r_0$ , where $b = 3 * 10^{-12}\mathrm{m}$, $v_0 = 1.59 * 10^7\mathrm{m/s}$ （corresponding to $b_0 = 1 * 10^{-12}\mathrm{m}$, $T = 1.11 * 10^7\mathrm{K}$） .

Secondly, the effect of incident distance on scattering angle is considered. The derivative of Eq. (29) for incident distance can be derived as

$$\frac{d\theta}{dr_0} \approx -\frac{2bb_0^2}{(b^2+b_0^2)r_0^2} - \frac{\sqrt{\frac{2eBb}{mv_0}}}{\left((b^2+b_0^2)\right)^{1/2}} - \frac{bb_0(8b^2+3b_0^2)+2b(b^2+b_0^2)r_0}{2r_0^2(2b_0+r_0)^2\sqrt{b^2+\frac{b_0^2 r_0}{2b_0+r_0}}} < 0. \tag{32}$$

It is shown that the scattering angle change with different incident distance. In our physical picture, the incident distance represents the influence range of magnetic field. Hence, the influence of magnetic field on scattering increases as incident distance so that the scattering angle will become smaller.

As shown in Fig. 3, the scattering angle decreases significantly as incident distance increases for different magnetic field and $b = 3 * 10^{-12}\mathrm{m}$, $v_0 = 1.59 * 10^7\mathrm{m/s}$.

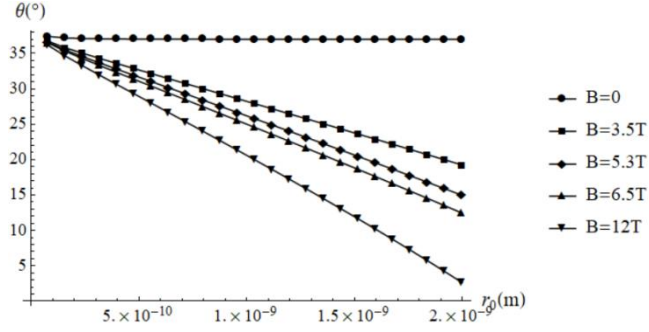

**FIG. 3.** The scattering angle decreases as $r_0$ increases in different B , where $b = 3 * 10^{-12}$m, $v_0 = 1.59 * 10^7$m/s.

Thirdly, the effect of the initial velocity on the scattering angle is also considered. The derivative of Eq. (29) for incident velocity can be obtained as

$$\frac{d\theta}{dv_0} \approx \frac{1}{v_0}\left(\frac{4bb_0(b_0 + r_0)}{\sqrt{1 + \frac{2b_0}{r_0}}(b_0^2 r_0 + 2b_0 b^2 + b^2 r_0)} + \frac{b\sqrt{-\epsilon}(b^2 - 3b_0^2)}{\sqrt{2}(b^2 + b_0^2)^{3/2}}\right) < 0 \qquad (33)$$

As shown in Fig. 4, the scattering angle decreases significantly as incident velocity increases for different magnetic field where $b = 5 * 10^{-11}$m, $r_0 = 1 * 10^{-9}$m. The reason is that the interaction time between electron and ion becomes shorter as incident kinetic energy is increasing. It also happens in Rutherford scattering for B=0.

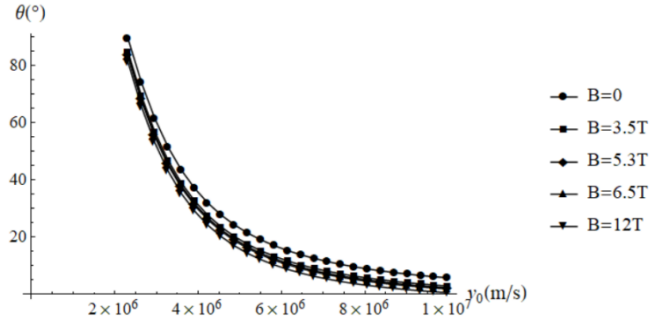

**FIG. 4.** The scattering angle decreases as $v_0$ increases in different B , where $b = 5 * 10^{-11}$m, $r_0 = 1 * 10^{-9}$m.

We use Mathematica to calculate the dependence of the scattering angle on the external magnetic field under specific conditions in order to compare our theoretical result with the numerical simulation.

The discrete form of the trajectory equation has been obtained by numerical calculation from equation (19). In fact, it is obtained by cubic spline interpolation. Then, combining this trajectory equation with the asymptote equation (7), the value of scattering angle in the case with/without external magnetic field is obtained, as shown in Fig. 5.

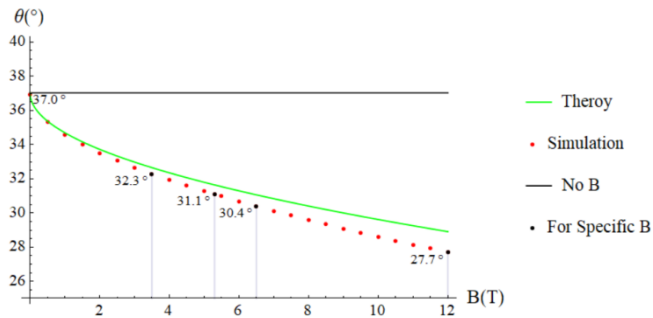

**FIG. 5.** The scattering angle in different magnetic field (B) both in theory (green line) and simulation (red and black points), where $r_0 = 5 * 10^{-10}$m, $b = 3 * 10^{-12}$m, $v_0 = 1.59 * 10^7$m/s

The scattering angle decreases as magnetic field is increasing. The difference of scattering angle with/without magnetic field is also shown in Fig. 5. We take $r_0 = 5 * 10^{-10}$m, $b = 3 * 10^{-12}$m, $b_0 = 1 * 10^{-12}$m, $v_0 = 1.59 * 10^7$m/s.

Under the selected parameters, the scattering angle is 37.0°, 32.3°, 31.1°, 30.4°, 27.7° respectively when the magnetic field is 0, 3.5T (EAST), 5.3T (ITER), 6.5T (CFETR), 12T (SPARC). The difference between the scattering angles with/without magnetic field is 14.7%, 19.0%, 21.3%, 25.2% respectively for 3.5T, 5.3T, 6.5T, 12T. Obviously, the influence of magnetic field on scattering angle cannot be ignored.

It can be seen that the theoretical approximation agrees well with the numerical simulation in weak magnetic field. The gap between the theoretical result and the numerical simulation increases as the magnetic field. The difference is 1.1%, 1.7%, 3.0%, 4.2% respectively when the magnetic field is 3.5T, 5.3T, 6.5T, 12T.

The Larmor radius of electron cyclotron is large for the weak magnetic field. For example, $\rho_e$ is about $9.0 * 10^{-6}m$ even when B = 10T. In our study, the physical picture of Rutherford scattering is still used. Therefore, the cyclotron motion is taken as the perturbation to the Coulomb collision. Rutherford scattering and cyclotron motion can be superimposed as the Larmor radius is much larger than $r_0$ ($\sim 5 * 10^{-10}$m), which just fits our hypothesis.

## V. CONCLUSION AND DISCUSSION

The purpose of this paper is to derive the analytical expression of scattering angle of electron-ion collision in the presence of an external magnetic field. For simplicity, the electron-ion Coulomb collisions perpendicular to the applied magnetic field are studied in this paper. The scattering angle is defined by the intersection of the electron trajectory under the external magnetic field and the asymptote of the electron trajectory without magnetic field. The asymptote equation of electron trajectory at finite incident distance without magnetic field is given. Meanwhile, the electron trajectory equation in an external magnetic field is derived. At present, a simple expression of scattering angle can be obtained under the weak magnetic field approximation. In the case of the external magnetic field, the electron-ion Coulomb scattering process is actually the superposition of electron cyclotron motion and Coulomb collision without magnetic field.

When the deflection caused by cyclotron motion partially offsets the Rutherford scattering angle, the electron trajectory will form a cross point with the asymptote without magnetic field. It is a reasonable to define the new scattering angle with the external magnetic field by using the position of the cross point in the collision center coordinate. The scattering angle is defined as the angle between the direction of the initial velocity of the electron and the direction of the radius vector of the cross point.

Because the exact analytical solution is very difficult to obtain, we take the cyclotron motion as the perturbation of the electron-ion Coulomb scattering process in the case of relatively weak magnetic field.

It is found that the scattering angle decreases significantly as external magnetic field increases for different incident distance. Obviously, the reason is that the influence of cyclotron motion on Coulomb collision becomes stronger as the external magnetic field increasing.

It is shown that the scattering angle decreases dramatically as incident distance increases for different external magnetic field because that the Coulomb interaction range becomes larger as the incident distance is increasing.

It is indicated that the scattering angle decreases observably as incident velocity increases for different magnetic field due to that the Coulomb interaction time becomes shorter as incident kinetic energy increasing.

The relation between scattering angle and external magnetic field has been calculated by using Mathematica in order to compare the theoretical result with numerical calculation.

It is found that the scattering angles decrease 14.7%, 19.0%, 21.3%, 25.2% respectively for 3.5T, 5.3T, 6.5T, 12T comparing with the case without magnetic field according to the numerical calculation.

It is indicated that the difference between the theoretical result and the numerical simulation is no more than 4.2% when $B \leq 12T$ as shown in Fig. 5.

In this paper, we only consider the case that the incident velocity is perpendicular to the magnetic field. We must consider the case that the incident velocity is not only perpendicular to the magnetic field if we want to derive the scattering cross section. This will be the next step of our work. Here we make an estimate for the isotropic case.

In Rutherford scattering process, the definition of differential collision cross section is

$$dN = Ibdbd\varphi = I\sigma d\Omega$$

$$\sigma = \frac{dN}{Id\Omega} = \frac{db}{d\theta} \frac{b}{\sin \theta}. \tag{34}$$

If only keeping the order of $\sqrt{\epsilon}$, we have

$$\frac{db}{d\theta} = \left(\frac{d\theta}{db}\right)^{-1} = \frac{\sqrt{2}(2b^2 b_0 + b^2 r_0 + b_0^2 r_0)(b^2 + b_0^2)^{1/2}}{-2\sqrt{2}b_0(b^2 + b_0^2)^{1/2}\sqrt{1 + \frac{2b_0}{r_0}}r_0 + r_0(b^2 - b_0^2)\sqrt{-\epsilon}} \tag{35}$$

and

$$\sin \theta = -\frac{2b_0 b\sqrt{\left(1 + \frac{2b_0}{r_0}\right)}}{b_0^2 + b^2\left(1 + \frac{2b_0}{r_0}\right)} - \frac{b\sqrt{-2\epsilon}}{\sqrt{(b^2 + b_0^2)\left(1 + \frac{2b_0}{r_0}\right)}} \frac{b^2 - b_0^2}{b_0^2 + b^2}. \tag{36}$$

Substituting Eqs. (35) and (36) into Eq. (34), we can obtain the expression for differential cross section as

$$\sigma = \frac{(b^2 + b_0^2)^2}{4b_0^2} + \frac{(b^4 - b_0^4)}{2b_0 r_0} + \frac{\sqrt{2}(b^2 + 3b_0^2)(b^2 + b_0^2)^{3/2}}{16b_0^3}\sqrt{\frac{eB\cos\psi\, r_0^2}{mbv_0}}. \tag{37}$$

When $B = 0$, $r_0 \to \infty$, the Equation (36) returns to

$$\sigma = \frac{(b_0^2 + b^2)^2}{4b_0^2} = \frac{b_0^2}{4\sin^4\frac{\theta}{2}}$$

with

$$\sin\frac{\theta}{2} = \frac{b_0}{\sqrt{b_0^2 + b^2}}.$$

It can be seen that the scattering cross section of Coulomb collision increases as applied magnetic field. This may be because the external magnetic field reduces the scattering angle of electrons, so that more electrons can fall into the scattering cross section.

Clearly, much more work is required to find scattering angle and the scattering cross section in three-dimensional case with external magnetic field.

## ACKNOWLEDGMENTS

It is grateful to Ms. Xue Lv for her previous work. This work was supported by the National MCF Energy R&D Program under Grant No. 2018YFE0311300, the National Natural Science Foundation of

China under Grant Nos. 11875067, 11835016, 11705275, 11675257, and 11675256, the Strategic Priority Research Program of Chinese Academy of Sciences under Grant No. XDB16010300, the Key Research Program of Frontier Science of Chinese Academy of Sciences under Grant No. QYZDJ-SSW-SYS016, and the External Cooperation Program of Chinese Academy of Sciences under Grant No. 112111KYSB20160039.

https://doi.org/10.1098/rspa.1992.0046